\documentclass[prc,aps,superscriptaddress,twocolumn,showpacs,nofootinbib]{revtex4}

\usepackage{graphicx,amsmath,amssymb,bm,multirow}
\usepackage{amsthm,mathrsfs}
\usepackage{amsfonts}
\usepackage{dcolumn}
\usepackage{bm}
\usepackage[colorlinks,linkcolor=blue,anchorcolor=blue,citecolor=blue]{hyperref}
\usepackage[usenames,dvipsnames]{color}
\newcommand{\beq}{\begin{equation}}
\newcommand{\eeq}{\end{equation}}
\newcommand{\beqn}{\begin{eqnarray}}
\newcommand{\eeqn}{\end{eqnarray}}

\newcommand{\bsub}{  \begin{subequations}}
\newcommand{\esub}{ \end{subequations}}

\begin{document}

%\title{Relativistic generator coordinate method for hypernuclear spectroscopy}
\title{Generator coordinate method for hypernuclear spectroscopy 
with a covariant density functional}

\author{H. Mei}
\affiliation{Department of Physics, Tohoku University, 
Sendai 980-8578,Japan}
\affiliation{School of Physical Science and Technology,
             Southwest University, Chongqing 400715, China}

\author{K. Hagino}
\affiliation{Department of Physics, Tohoku University, 
Sendai 980-8578,Japan}
\affiliation{Research Center for Electron Photon Science, 
Tohoku University, 1-2-1 Mikamine, Sendai 982-0826, Japan}
\affiliation{
National Astronomical Observatory of Japan, 2-21-1 Osawa,
Mitaka, Tokyo 181-8588, Japan}

\author{J. M. Yao 
\footnote{Present address: Department of Physics and Astronomy, University of North Carolina, 
Chapel Hill, North Carolina 27516-3255, USA}
}
\affiliation{Department of Physics, Tohoku University, 
Sendai 980-8578,Japan}
\affiliation{School of Physical Science and Technology,
             Southwest University, Chongqing 400715, China}

\begin{abstract}
We apply the generator coordinate method (GCM) 
to single-$\Lambda$ hypernuclei 
in order to discuss 
the spectra of hypernuclear low-lying states. 
To this end, we 
use the same relativistic point-coupling energy functional 
both for the mean-field and the beyond-mean-field calculations. 
This relativistic GCM approach provides a unified description 
of low-lying states in ordinary nuclei and in hypernuclei, 
and is thus suitable for studying the $\Lambda$ impurity 
effect. 
We carry out 
an illustrative calculation for the low-lying spectrum 
of $^{21}_\Lambda$Ne, in which the interplay between the hypernuclear 
collective excitations and the single-particle excitations 
of the unpaired $\Lambda$ hyperon is 
taken into account in a full microscopic manner. 
\end{abstract}

%21.80.+a	Hypernuclei
%21.60.Jz	Nuclear Density Functional Theory and extensions (includes Hartree-Fock and random-phase approximations)
%23.20.-g   Electromagnetic transitions
%21.10.Ky	Electromagnetic moments
%27.30.+t	20 ¡Ü A ¡Ü 38
%27.20.+n	6 < A < 19
%21.10.-k   Properties of nuclei; nuclear energy levels
\maketitle

%\section{Introduction}
In the past decades, many high-resolution $\gamma$-ray 
spectroscopy experiments have been carried out for 
$\Lambda$ hypernuclei. The experimental data on 
energy spectra and electric multipole transition strengths 
have been accumulated, providing rich information on a 
hyperon-nucleon interaction in the nuclear medium as well as the 
impurity effect of the $\Lambda$ particle 
on the structure of atomic nuclei~\cite{Hashimoto06,Tamura09}. 
It is noteworthy that the next-generation facility J-PARC has 
already been in operation \cite{Yamamoto}, opening up a 
new opportunity to 
perform high precision 
hypernuclear 
$\gamma$-ray spectroscopy studies. 
These experiments will shed light on 
low-lying states of hypernuclei, 
especially those of medium and heavy hypernuclei.

From the theoretical side, the hypernuclear low-lying states 
have been studied mainly with a shell 
model~\cite{Dalitz78,Gal71,Millener} and with a cluster and 
few-body models~\cite{Motoba83,Hiyama99,Bando90,Hiyama03,Cravo02,
Suslov04,Shoeb09}. In recent years, several other methods have 
also been developed for hypernuclear spectroscopy, including 
an ab-initio method~\cite{abinitio}, the antisymmetrized 
molecular dynamics (AMD)~\cite{Isaka11,Isaka12,Isaka13}, and 
the microscopic particle-rotor model based on the 
covariant density functional theory~\cite{Mei2014,Mei2015}.
The angular momentum projection (but with the scheme of 
variation-before-projection) for the total hypernuclear 
wave function has also been carried out with the Skyrme density 
functional \cite{Cui15}, even though the important 
effect of configuration 
mixing was not taken into account. 

In this paper, 
we propose a generator coordinate method (GCM) 
for the whole hypernucleus based on a relativistic energy 
density functional.  
To this end, we superpose a set of hypernuclear mean-field 
states projected onto the states with good quantum numbers of 
the particle number and the angular momentum. 
Such configuration mixing effect was missing in 
Ref. \cite{Cui15}, and thus our calculation 
serves as one of the most advanced beyond-mean-field methods 
for the spectroscopy of hypernuclear low-lying states.
In contrast to the microscopic particle-rotor model developed in 
Refs. \cite{Mei2014,Mei2015}, 
where the GCM calculation is carried out only for the core 
nucleus, 
all the nucleons and the hyperon are treated on the same 
footing in the present approach. As we shall discuss, 
these two methods are in fact complementary 
to each other, both from the physics point of view and from the 
numerical point of view. 

Our aim in this paper is to describe 
low-lying states of odd-mass $\Lambda$ hypernuclei 
which consist of a $\Lambda$ particle and an even-even nuclear core. 
In contrast to the unpaired nucleon in ordinary odd-mass nuclei, 
the unpaired $\Lambda$ hyperon in the hypernucleus is free from the 
Pauli exclusion principle from the nucleons inside the nuclear core. 
In principle, the 
$\Lambda$ hyperon can thus occupy any bound hyperon orbital, 
providing 
an unique platform to study the interplay of the individual 
single-particle motion of the hyperon with the 
nuclear collective motions. 
The wave function of these hypernuclear states are constructed as 
a superposition of quantum-number projected hypernuclear 
reference states 
with different quadrupole deformation $\beta$,
\beq
\label{GCM:wf}
| \Psi^{JM}_{n\alpha} \rangle
= \sum _{\beta} f^{J}_{n\alpha}(\beta)\hat{P}^J_{MK}\hat{P}^N \hat{P}^Z|\Phi^{(N\Lambda)}_{n}(\beta)\rangle,
\eeq
where 
the index $n$ refers to a different hyperon orbital state, 
and the index $\alpha$ labels the quantum numbers of the state 
other than the angular momentum. 
For simplicity, we take the adiabatic approximation and 
do not mix different $n$ in the total wave function,  
$| \Psi^{JM}_{n\alpha} \rangle$.

In Eq. (\ref{GCM:wf}), 
the mean-field states $|\Phi^{(N\Lambda)}_{n}(\beta)\rangle$, 
severing as nonorthonormal basis, are generated with 
deformation constrained relativistic mean-field (RMF) 
calculations for $\Lambda$ hypernuclei~\cite{Myaing08,Lu11,Weixia15}. 
Since the hyperon and the nucleons are not mixed, 
the mean-field states can be decomposed as
\beq
|\Phi^{(N\Lambda)}_{n}(\beta)\rangle
= |\Phi^N(\beta)\rangle \otimes |\varphi^{\Lambda}_{n}(\beta)\rangle,
\eeq
where $|\Phi^N(\beta)\rangle $ and 
$|\varphi^{\Lambda}_{n}(\beta)\rangle$ are the mean-field wave 
functions for the nuclear core and the hyperon, respectively.
With this wave function, 
the deformation parameter $\beta$ is related to the mass 
quadrupole moment of the whole hypernucleus $^A_\Lambda Z$ as 
\beq
\beta = \dfrac{4\pi}{3AR^2}\langle\Phi^{(N\Lambda)}_{n}(\beta)
\vert r^2 Y_{20} \vert \Phi^{(N\Lambda)}_{n}(\beta)\rangle,
\eeq
with $R=1.2\times A_c^{1/3}$ fm, $A_c=A-1$ being the mass 
number of the core nucleus. 
In this paper, in order to 
reduce the computation burden, we restrict 
all the reference states to be axially deformed. 
The mean-field states $|\Phi^{(N\Lambda)}_{n}(\beta)\rangle$ are 
then projected onto states with good quantum numbers with 
the operators 
$\hat{P}^{N}$ ($\hat{P}^{Z}$), and $\hat P^J_{MK}$, 
which project out the component with good neutron (proton) 
numbers and the angular momentum~\cite{Ring80}.
Here, the total angular momentum $J$ is a half-integer 
number and 
$K$ is its projection on the $z$-axis in the body-fixed frame. 
We assume that all the nucleons fill time-reversal states 
and thus do not contribute to the total angular momentum 
along the symmetric axis. In this case, 
the $K$ quantum number 
is identical to $\Omega$, that is, the component 
of the angular momentum of the hyperon along the $z$-axis, 
and thus can be adopted to characterize the wave function 
$|\varphi^{\Lambda}_{n}(\beta)\rangle$. 
From the mean-field states 
with the hyperon in a $\Omega^\pi$ 
configuration, 
the angular momentum $J$ takes the value of 
$\vert \Omega\vert, \vert \Omega\vert+1, \cdots$. 
Notice that, in the angular momentum projection, the integrals 
over the two Euler angles $\phi$ and $\psi$ can be 
performed analytically because of the axial symmetry. 

The weight function $f^{J }_{n\alpha}(\beta)$ in the GCM states 
given by Eq. (\ref{GCM:wf}) 
is 
determined by the variational principle, which leads to the 
Hill-Wheeler-Griffin (HWG) equation, 
 \beq
 \label{HWE}
 \sum_{\beta'}
 \left[{\cal H}^J_{n}(\beta,\beta') -E^{J}_{n\alpha}
{\cal N}^J_{n }(\beta,\beta')\right]
  f^{J}_{n\alpha}(\beta')=0,
 \eeq
 where the norm kernel 
${\cal N}^J_{n }(\beta,\beta')$ 
and the Hamiltonian kernel 
${\cal H}^J_{n }(\beta,\beta')$ 
are defined as
\beq
{\cal O}^J_{n}(\beta,\beta')\equiv
 \langle \Phi^{(N\Lambda)}_{n}(\beta) \vert \hat O
\hat{P}^J_{KK}\hat{P}^N \hat{P}^Z \vert\Phi^{(N\Lambda)}_{n}
(\beta')\rangle,
\eeq
with $\hat{O}=1$ and $\hat{O}=\hat{H}$, respectively. 
The solution of the HWG equation (\ref{HWE}) provides the energy 
$E^{J}_{n\alpha}$ 
and the weight function $f^{J}_{n\alpha}(\beta)$ 
for each of the low-lying states of hypernuclei.
In the actual calculations, 
we evaluate the Hamiltonian overlap with the mixed 
density prescription~\cite{Yao09,Yao10}. 

\begin{figure}[]
  \centering
 \includegraphics[width=7.5cm]{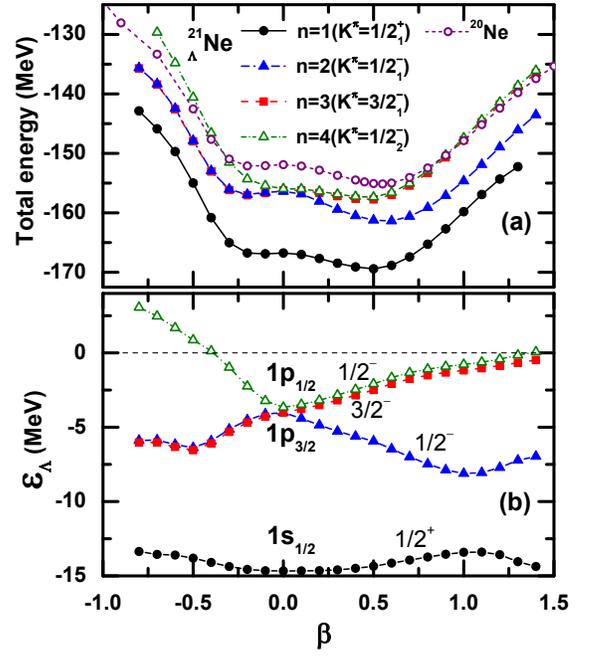}
 \caption{(Color online)(a) The total energy curves 
for $^{21}_\Lambda$Ne obtained in the mean-field approximation 
as a function of quadrupole deformation $\beta$.  
These are calculated by putting 
the $\Lambda$ hyperon in different single-particle orbitals 
shown in the lower panel.
For comparison, 
the energy curve for the core nucleus $^{20}$Ne is also 
plotted. 
 (b) The single-particle energies of the 
$\Lambda$ hyperon in  $^{21}_\Lambda$Ne as a 
function of quadrupole deformation. 
These are labeled with 
the $\Omega^\pi$ number, that is 
the projection of the angular momentum 
onto the $z$-axis in the body fixed frame. }
\label{MFcurve}
\end{figure}

As an illustration of the method, 
we apply the GCM approach to $^{21}_\Lambda$Ne. 
We first generate a set of hypernuclear reference states 
$|\Phi^{(N\Lambda)}_{n}(\beta)\rangle$, by putting the hyperon 
on the four lowest single-particle states with 
$\Omega^\pi=1/2^+_1, 1/2^-_1, 3/2^-_1$, and $1/2^-_2$. 
To this end, we perform 
the deformation constrained RMF+BCS calculation 
using the PC-F1 force~\cite{Buvenich02} for the nucleon-nucleon 
interaction and the PCY-S2 force~\cite{Tanimura2012} for the 
nucleon-$\Lambda$ interaction. 
A density-independent $\delta$ force 
is used in the pairing channel for the nucleons, supplemented 
with an energy-dependent cutoff~\cite{Bender00}. 
The Dirac equations are solved by expanding the Dirac spinors 
with harmonic oscillator wave functions with 10 oscillator shells. 
See Refs.~\cite{Weixia15,Yao14} for numerical details. 

Figure~\ref{MFcurve}(a) shows the mean-field energies for the 
reference states so obtained as a function of 
deformation parameter $\beta$. 
One can see that 
the energies for the three negative-parity 
configurations (that is, $\Omega^\pi=1/2^-_1, 3/2^-_1$, and $1/2^-_2$), 
corresponding to the hyperon occupying the 
three ``$p$-orbital" states, are close 
to each other at $\beta=0$ due to a weak hyperon spin-orbit 
interaction, and are well separated from the energy of the 
positive parity configuration ($\Omega^\pi=1/2^+_1$), which 
corresponds to the hyperon occupying the ``$s$-orbital" state. 
The energy difference between the positive- and the 
negative-parity energy configurations at $\beta=0$ 
is about 10.4 MeV, which is 
consistent with the $2/3$ of the energy scale 
$\hbar\omega=41A^{-1/3}$ MeV for nucleons. 
This energy 
corresponds to the excitation energy of hyperon from the 
$s$-orbital to the $p$-orbital.
Moreover, one can also see that 
the energy minimum appears at 
$\beta\sim0.6$ for $K^\pi=1/2^-_1$, 
which is larger 
than the deformation of the energy minimum 
for the $1/2^+_1$ configuration ($\beta=0.49$). 
This is consistent with 
the findings in Refs.~\cite{Isaka11,Weixia15} that the hyperon 
in the ``$p$-orbital" tends to develop a pronounced energy minima 
with a larger deformation.

Figure~\ref{MFcurve}(b) shows the Nilsson diagram for the 
hyperon in $^{21}_\Lambda$Ne. The single-particle level with 
the $\Omega^\pi=3/2^-_1$ configuration 
is approximately degenerate with 
the $\Omega^\pi=1/2^-_1$ and $1/2^-_2$ configurations 
at the oblate and the prolate sides, respectively. 
This is a characteristic feature of the Nilsson diagram without 
the spin-orbit interaction \cite{Ring80}, 
and is responsible for the approximate degeneracy of the 
corresponding total energy curves shown in Fig.~\ref{MFcurve}(a). 
We note that the second $1/2^-$ single-particle level becomes 
unbound on the oblate side with deformation parameter of $\beta<-0.3$. 
In the following discussions, we therefore focus on 
the hypernuclear states generated by the $\Lambda$ hyperon occupying 
the $\Omega^{\pi}=1/2^+_1,1/2^-_1$, and $3/2^-_1$ configurations. 

\begin{figure}[]
  \centering
 \includegraphics[width=8.5cm]{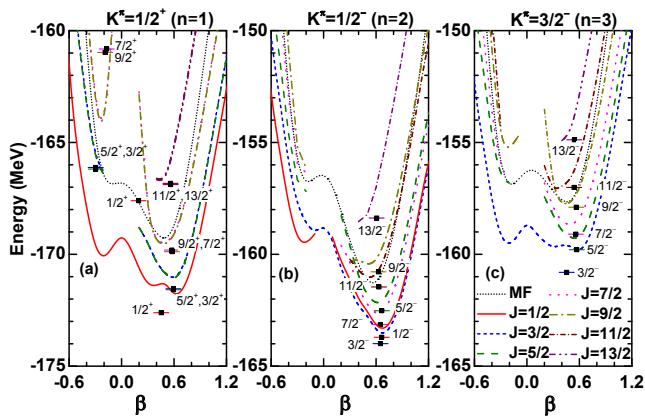}
 \caption{(Color online) The projected energy curves for 
$^{21}_\Lambda$Ne obtained by putting  
the $\Lambda$ hyperon on the three lowest single-particle orbitals 
labeled by $\Omega^\pi(=K^\pi)$. 
The corresponding mean-field energy curves 
are also shown for a comparison. 
The solutions of the GCM calculations 
are indicated 
by the squares and the horizontal bars 
placed at the average deformation.}
\label{ProjCurve}
\end{figure}

The energy curves shown in 
Fig.~\ref{MFcurve}(a) are the results of the mean-field 
approximation, in which 
the reference states are not the eigen-states of 
the angular momentum and the nucleon numbers. 
The projected energy curves, after the projection procedures, 
are obtained by taking the 
diagonal element of the Hamiltonian and the norm kernels 
as $E^J_n(\beta)=
{\cal H}^J_{n}(\beta,\beta)/{\cal N}^J_{n}(\beta,\beta)$. 
Those energy curves are 
plotted in Fig.~\ref{ProjCurve} as a function of $\beta$.  
For the $K^\pi=1/2^+_1$ configuration shown in 
Fig.~\ref{ProjCurve}(a), the projected energy 
curves for $J^\pi=3/2^+$ and $5/2^+$ 
almost overlap with each other, 
indicating a weak coupling of the $\Lambda$ hyperon to the 
nuclear core. This is the case also for the pairs of 
$J^\pi=(7/2^+, 9/2^+)$ and 
$J^\pi=(11/2^+, 13/2^+)$. 
It is seen that 
the prolate 
minimum in the projected energy curves becomes more pronounced 
and thus the nuclear shape becomes more stable as 
the angular momentum increases. 
Moreover,  the energy minimum for the 
$J^\pi=1/2^+$ energy curve appears at 
deformation $\beta=0.62$, that is somewhat larger than 
the deformation at the minimum of 
the corresponding mean-field curve, $\beta=0.49$,  
due to the energy 
gain originated from the angular momentum projection.
On the other hand, 
if one compares it to the projected energy curve for the 0$^+$ 
configuration of $^{20}$Ne, which has a minimum at $\beta=0.65$,   
one finds that the minimum 
is slightly shifted towards the spherical 
configuration both on the oblate and the prolate sides, 
similarly to 
the finding of the microscopic particle rotor 
model~\cite{Mei2015}. 

In contrast to the $J^\pi=1/2^+$ configuration, 
the deformation at 
the energy minimum for the 
$J^\pi=1/2^-$ configuration increases 
to $\beta=0.69$ (see 
Fig.~\ref{ProjCurve}(b)). 
Moreover, for this configuration, the energy difference 
between the prolate and the oblate minima significantly increases 
as compared to 
the $J^\pi=1/2^+$ configuration. 
For this reason, the collective wave function for the $J^\pi=1/2^-$ 
state is expected to be more localized on the prolate side 
than that of the $J^\pi=1/2^+$ state. 
As a consequence, the average deformation for the $J^\pi=1/2^-$ 
state is close to the minimum point of the energy curve 
while that for the $J^\pi=1/2^+$ configuration is shifted towards 
the oblate side due to a cancellation between the prolate and the 
oblate contributions (see the filled squares 
in Fig.~\ref{ProjCurve}(a) 
and \ref{ProjCurve}(b)).  

The projected energy curves for the $K^\pi=3/2^-_1$  
configuration are shown in Fig.~\ref{ProjCurve}(c).
These are 
several MeV higher than those for the $K^\pi=1/2^-_1$ 
configuration. Besides, 
the energy curve for the $J^\pi=3/2^-$ is considerably different 
from that for the $J^\pi=5/2^-$ configuration, and one would not 
expect a (quasi-)degeneracy between these two states. 

\begin{figure}[t]
  \centering
 \includegraphics[width=8.5cm]{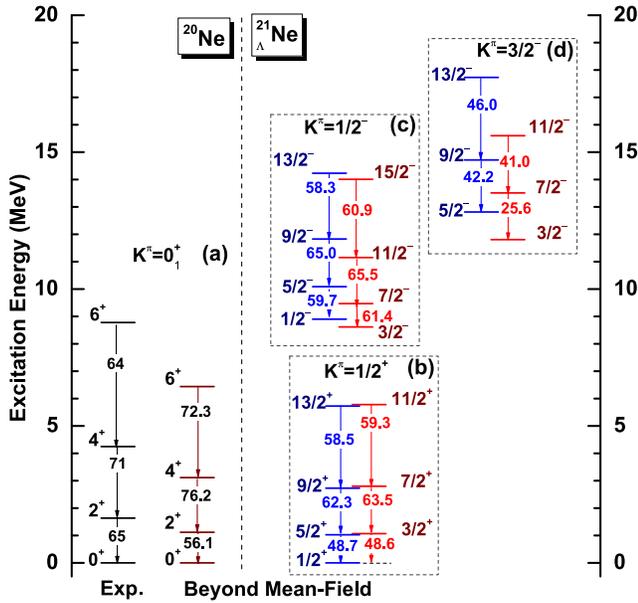}
 \caption{(Color online) The low-lying excitation spectra of 
$^{20}$Ne (a) and  $^{21}_\Lambda$Ne [(b)-(d)] constructed with 
the GCM approach. 
The numbers with the arrows indicate the $E2$ transition 
strengths, given in units of $e^2$ fm$^4$. 
The experimental data for $^{20}$Ne are 
taken from Ref.~\cite{lbl}.
}
\label{spectra}
\end{figure}

By mixing all the projected mean-field states for each $K^\pi$ 
configuration, we 
construct the low-lying states of $^{20}$Ne and $^{21}_\Lambda$Ne 
with the GCM method. 
The calculated spectra are shown in 
Fig.~\ref{spectra}. 
One can see that 
the rotational character of the yrast states of $^{20}$Ne 
is well reproduced, although the moment of inertia is somewhat 
overestimated due to the pairing collapse in the reference states 
for $\beta>0.5$. 
This problem is expected to be improved by introducing 
the method of particle-number projection before variation 
while generating the reference states. 
The $\Lambda$ binding energy of $^{21}_\Lambda$Ne, 
defined as the energy difference between the $0^+_1$ state 
of $^{20}$Ne and the $1/2^+_1$ state of $^{21}_\Lambda$Ne, 
is calculated to be $B_{\Lambda}=14.11$ MeV, which is 
slightly smaller than the mean-field result of 14.27 MeV.

According to a naive picture of a deformed rotor coupled to 
a hyperon moving in the deformed potential, 
one may expect several rotational bands 
with angular momenta in the order of 
$J=\vert \Omega\vert, \vert \Omega\vert+1, \cdots$ 
built on top of 
each single-particle state of $\Lambda$ hyperon with 
$\Omega^\pi$. 
This picture is indeed realized for the $K^\pi=3/2^-$ band 
shown in Fig.~\ref{spectra} (d), but is somewhat distorted for 
the $K^\pi=1/2^+$ (b) and $1/2^-$ (c) bands. 
In particular, the spin-parity of the bandhead state of 
the $K^\pi=1/2^-$ band is not $J^\pi=1/2^-$, but $J^\pi=3/2^-$ 
due to a large decoupling factor originated from the 
Coriolis interaction \cite{Ring80}. 
This effect inverts the energy ordering of the states in 
the $K^\pi=1/2^-$ band by shifting up the states with odd value 
of $J+1/2$ 
and pulling down the states with even values of $J+1/2$. 
As a result, two rotational bands having $\Delta J=2$ and 
with similar electric quadrupole transition strengths are formed.  
A similar feature has also been found in the microscopic 
particle-rotor model calculation~\cite{Mei2015}, where 
the energy displacement between the two bands is, however, 
much smaller. 
To be more specific, 
the energy difference between the $1/2^-$ and $3/2^-$ states 
is less than 40 keV with the microscopic particle-rotor model, while 
it is 270 keV with the present GCM calculation. 
We have confirmed that this feature remains the same even 
if we mix the 
$\Omega^\pi=1/2^-_1$ and $1/2^-_2$ configurations in the 
GCM calculations, which alters the excitation energies only 
by $\sim$2$\%$. 

The $K^\pi=1/2^+$ band is mainly formed by the 
$\Lambda$ hyperon in the ``$s$-orbital"  
coupled to the ground-state band 
of the nuclear core, $^{20}$Ne. 
For each core state, except for the ground state,  
two states appear in this 
band due to the angular momentum coupling with 
$j=1/2$, and two rotational series are formed. 
The energy splitting in the double states is predicted to be 
small. 
That is, it is 
41.5 keV, 71.2 keV and 53.8 keV, for the doublets  
$(3/2^+, 5/2^+)$, $(7/2^+, 9/2^+)$ and $(11/2^+, 13/2^+)$, 
respectively. 
The magnitude of these energy splittings is comparable to 
the empirical energy splitting of $^9_{\Lambda}$Be, for which 
the energy of the $5/2^+$ state is lower than the energy of 
the state $3/2^+$ by 43 keV~\cite{Tamura05}. 
For the E2 transition strength for $3/2^+ \rightarrow 1/2^+$ in 
$^{21}_\Lambda$Ne, we find that it 
is smaller than the E2 strength for 
$2^+ \rightarrow 0^+$ in $^{20}$Ne 
by $13.37\%$. 
This implies that the $\Lambda$ hyperon in the ``$s$-orbital" 
decreases the quadrupole collectivity of $^{20}$Ne, 
which is consistent with the findings in recent 
theoretical studies~\cite{Isaka11,Weixia15,Mei2015,YLH11}. 
We notice that this is consistent also with the 
distribution of the collective wave functions, 
which are shifted towards the small deformation region 
as compared to those of $^{20}$Ne. 
On the other hand, 
the impurity effect for the $\Lambda$ hyperon in the ``$p$-orbital" 
is 
more difficult to assess, because 
several configurations are admixtured in the wave functions, 
as has been shown 
in Ref.~\cite{Mei2015} with 
the microscopic particle-rotor model. 

In summary, we have applied the generator coordinate method to 
the spectroscopy of hypernuclear low-lying states. This approach 
is based on the beyond-mean-field method with 
the particle number and the angular momentum projections, 
and takes into account 
the interplay between the 
single-particle motion of the $\Lambda$ hyperon 
and the 
hypernuclear collective motion. 
Using a 
relativistic point-coupling energy density functional, we have 
carried out a proof-of-principle calculation 
for the low-lying states of $^{21}_\Lambda$Ne. 
Our results indicate that the $\Lambda$ hyperon 
in the ``$s$-orbit" 
couples weakly to the ground-state rotational band of 
the core nucleus $^{20}$Ne, forming a similar rotational band 
with almost degenerate doublet states. The $E2$ transition 
strength for the $2^+ \rightarrow 0^+$ transition 
in $^{20}$Ne is reduced by $13.37\%$ by adding the $\Lambda$ 
particle. On the other hand, 
for the $\Lambda$ hyperon in the ``$p$-orbit" 
the energy difference between similar doublet states is much 
larger, although 
the rotational structure of the $^{20}$Ne is still preserved. 

We note that for low-lying states of hypernuclei 
the unpaired hyperon is mainly filled in the 
deformed states with relatively low orbital angular momenta (that is, 
$s$ and $p$-like states) and 
is treated separately in the present GCM approach 
from the other nucleons.  
Therefore, the numerical calculation is much simpler than the 
GCM calculations for ordinary odd-mass nuclei, which 
has recently been developed based on a Skyrme energy density 
functional~\cite{Bally14}. 
Moreover, it is straightforward to extend the present method 
to include more complicated reference states, such as those with 
triaxial 
and octupole deformations, even though 
the numerical complexity will rapidly increase. 
We note that $^{20}$Ne has prominent negative-parity bands 
originated from the $\alpha$+$^{16}$O cluster structure, which would also 
exist in $^{21}_\Lambda$Ne. 
It would be interesting to study 
how the octupole deformation in the mean-field states modifies 
the low-lying states of $^{21}_\Lambda$Ne. 

The GCM approach presented in this paper is complementary to the 
microscopic particle-rotor approach which we have developed in 
the earlier publications \cite{Mei2014,Mei2015}. 
The wave functions for hypernuclear states are expressed 
in different ways in these approaches. 
In the microscopic particle rotor model, 
hypernuclear states are expanded in terms of the low-lying states 
of the core nucleus, while they are generated from 
intrinsic states for the whole system in the present GCM approach. 
Both methods have advantages and disadvantages. 
In the microscopic particle-rotor model, the non-adiabatic effects of $\Lambda$ particle is 
automatically taken into account, while the $\Lambda$ particle is restricted to a specific 
single-particle configuration in the present 
GCM approach, although this restriction may be easily 
removed. Another point is that the cut-off of the nuclear core states has to be introduced 
in the particle-rotor model, while one does not need to worry about it in the GCM approach. 
From a physics point of view, 
the microscopic particle-rotor model provides a convenient way to analyze the components 
of hypernuclear wave function, while 
the GCM approach offers an intuitive way to study 
the hypernuclear shape fluctuation as well as 
the nuclear shape polarization due to the 
$\Lambda$ hyperon. 
From a numerical point of view, the GCM approach is 
numerically more expensive than the microscopic particle-rotor 
model, since the norm and the Hamiltonian kernels have to 
be constructed for each $J^\pi$, 
whereas 
it is sufficient to do it only for a limited 
number of the core states ({\it e.g.,} $J^\pi=0^+,2^+$, and $4^+$) 
in order to construct the whole hypernuclear spectrum with the 
miscroscopic particle-rotor model.
So far, the microscopic particle-rotor model calculations have been carried out only with a simplified 
nucleon-hyperon interaction. 
A comparative study of these two methods using the same nucleon-hyperon effective interaction 
is an interesting future project, which will deepen 
our understanding on hypernuclear spectroscopy.

%%==========================================================================
%\section*{Acknowledgments}
\bigskip

We thank Xian-Rong Zhou for useful discussions. 
This work was supported in part by the Tohoku University Focused 
Research Project \lq\lq Understanding the origins for matters in 
universe\rq\rq, JSPS KAKENHI Grant Number 2640263, the NSFC under 
Grant Nos. 11575148 and 11305134.
%%==========================================================================

\end{document}